# Title: Targeted marine cloud brightening can dampen El Niño

**Authors:** Jessica S. Wan[1], John T. Fasullo[2], Nan Rosenbloom[2], Chih-Chieh Jack Chen[2], Katharine Ricke[1,3]*

**Affiliations:**

[1]Scripps Institution of Oceanography, University of California San Diego, La Jolla, CA, USA

[2]Climate and Global Dynamics Laboratory, National Center for Atmospheric Research, Boulder, CO, USA

[3]School of Global Policy and Strategy, University of California San Diego, La Jolla, CA, USA

*Corresponding author. Email: kricke@ucsd.edu

**Abstract:** Many record-breaking climate extremes arise from both greenhouse gas-induced warming and natural climate variability. Marine cloud brightening, a solar geoengineering strategy originally proposed to reduce long-term warming, could potentially mitigate extreme events by instead targeting seasonal phenomena, such as El Niño-Southern Oscillation (ENSO). By exploiting the 2019-2020 Australian wildfire experiment-of-opportunity, we show that simulated marine cloud brightening in the southeast Pacific reproduces observed cloud changes and induces La Niña-like responses. We then explore how cloud brightening timing and duration modifies the 1997-1998 and 2015-2016 El Niño events. We find the earliest and longest interventions effectively restore neutral ENSO conditions and dampen El Niño's impacts. Solar geoengineering that targets climate variability could complement tools such as ENSO forecasting and provide a pathway for climate risk mitigation.

**Main Text:**

Many extreme weather events in the past decades have resulted from the compounding effects of short-term stochastic events, long-term anthropogenic responses from greenhouse gases, and interannual or seasonal natural climate variability (*1*, *2*). One such mode is El Niño-Southern Oscillation (ENSO), the most ubiquitous source of interannual climate variability (*3*) which dramatically influences extreme weather globally (*4*, *5*). Because global temperatures are higher than average during El Niño (weakened trade winds and upwelling, equatorial sea surface temperature [SST] warming), these events can amplify effects associated with gradual global warming (*2*, *6*). Even absent climate change, while the societal impacts of El Niño events are heterogeneous, on net, El Niño is costly to the global economy on the order of trillions of dollars (*7*). La Niña (enhanced trade winds and upwelling, equatorial SST cooling) events can have beneficial effects for some regions, but the changes are generally smaller and insignificant (*7–9*).

Solar geoengineering (SG), a set of approaches to increase the amount of sunlight reflected to space (*10*), was originally proposed as a way to mitigate the steady, long-term warming from greenhouse gas emissions. However, SG could theoretically be leveraged to mitigate extreme events by instead targeting compounding seasonal events such as El Niño. While there have been a few studies to analyze the effect of SG on ENSO strength and variability (*11–14*), none have been designed to specifically target ENSO through plausible regional deployment.

One SG proposal that could be amenable to targeted application is marine cloud brightening (MCB) (*15*, *16*), which was proposed as a way to cool the planet by injecting aerosols into the lower atmosphere to form brighter marine clouds. Global implementation of MCB can dampen ENSO variability (*17*), however brightening only the clouds in the Southeast Subtropical Pacific (SESP) region has been robustly linked to La Niña-like responses (*18–20*). Recent work also suggests MCB could be leveraged for regional climate or sociopolitical objectives (*21–24*) in addition to the global responses targeted by early studies (*18*, *19*, *25*, *26*).

While ship tracks have long been considered the closest observational analogue to MCB, the unprecedented 2019-2020 Australian wildfires present a novel “experiment-of-opportunity” for MCB. Using the Community Earth System Model version 2 (CESM2) (*27*), it has been shown that biomass burning aerosols emitted from the bushfires were transported across the South Pacific and brightened the SESP stratocumulus cloud deck, which triggered dynamical responses that contributed to the 2020-2023 multi-year La Niña (*28*). This observed enhancement in cloud albedo, lifetime, and extent serves as a natural analogue that can be used to understand how cloud modification, artificially or naturally induced, could play a role in modifying climate variability as well as the mean state.

In this study, we explore the feasibility, from a physical climate perspective, of implementing targeted MCB to deliberately modify ENSO. We first test our hypothesis that the 2019-2020 Australian wildfire event contributing to La Niña is a good analogue for how MCB might influence ENSO. We then investigate the viability of deploying MCB after the spring predictability barrier for historical El Niño events to deliberately weaken El Niño.

To model MCB’s influence on ENSO, we use a skilled seasonal initialized prediction tool (*29*) using CESM2 (*27*). We run a set of two-year simulations with ten ensemble members each, initialized prior to three characteristic historical ENSO events: (1) 2020-2021 La Niña, (2) 2015-2016 El Niño, and (3) 1997-1998 El Niño (table S1). MCB perturbations are modeled following the methodology in Wan et al. (2024) by nudging the cloud droplet number concentration to 500 particles/$cm^3$ over the SESP (see 'Simulating MCB with the Seasonal-to-Multiyear Large Ensemble using CESM2', in Materials and Methods). While there are many ways to represent MCB in Earth system models including fixed radiative forcing and aerosol emission, the resultant climate responses are more sensitive to the magnitude and location of perturbation than the representation of MCB in CESM2 (*20*). Our approach of achieving a cloud droplet number concentration of 500 particles/$cm^3$ over the SESP region (~5% of the Earth's surface) could hypothetically be accomplished with sprayers attached to roughly 2% of the world merchant fleet (*30*) using limited estimates from the literature (*31*), a substantial but technically feasible allocation of existing resources.

## Wildfires as a natural analogue for marine cloud brightening

To explore the viability of the 2019-2020 Australian wildfire event as a proxy for MCB, we compare the climate responses from the wildfires (*28*) (Figs. 1A-C) to simulations where instead MCB is deployed in the SESP region from January to March 2020 (Figs. 1D-F). This three-month MCB window was chosen to align with the period of enhanced cloud albedo in the SESP rather than the peak of the bushfire emissions which occurred in December 2019 to January 2020 (*28*).

We find that MCB reproduces many of the key mechanisms identified from the wildfires including an immediate negative local shortwave forcing response (Figs. 1A and 1D) leading to boundary layer drying and a northward shift of the Intertropical Convergence Zone (Figs. 1B and 1E), and subsequent cooling in the Niño3.4 region (*28*) (Figs. 1C and 1F). The January to March global mean shortwave cloud forcing is higher in our MCB experiment (-20.1 W/$m^2$) compared to the wildfires (-6.3 W/$m^2$), which leads to stronger global mean DJF cooling (114% increase). The temperature pattern from such intense cooling is not identical to the wildfire response, but potentially more akin to a multi-year La Niña with cooling in the central/western Pacific (*32*). Nonetheless, simulated MCB reproduces similar La Niña-like responses in the eastern Pacific to the 2019-2020 Australian wildfires, suggesting a plausible observational analogue for what could happen due to an MCB intervention in the SESP.

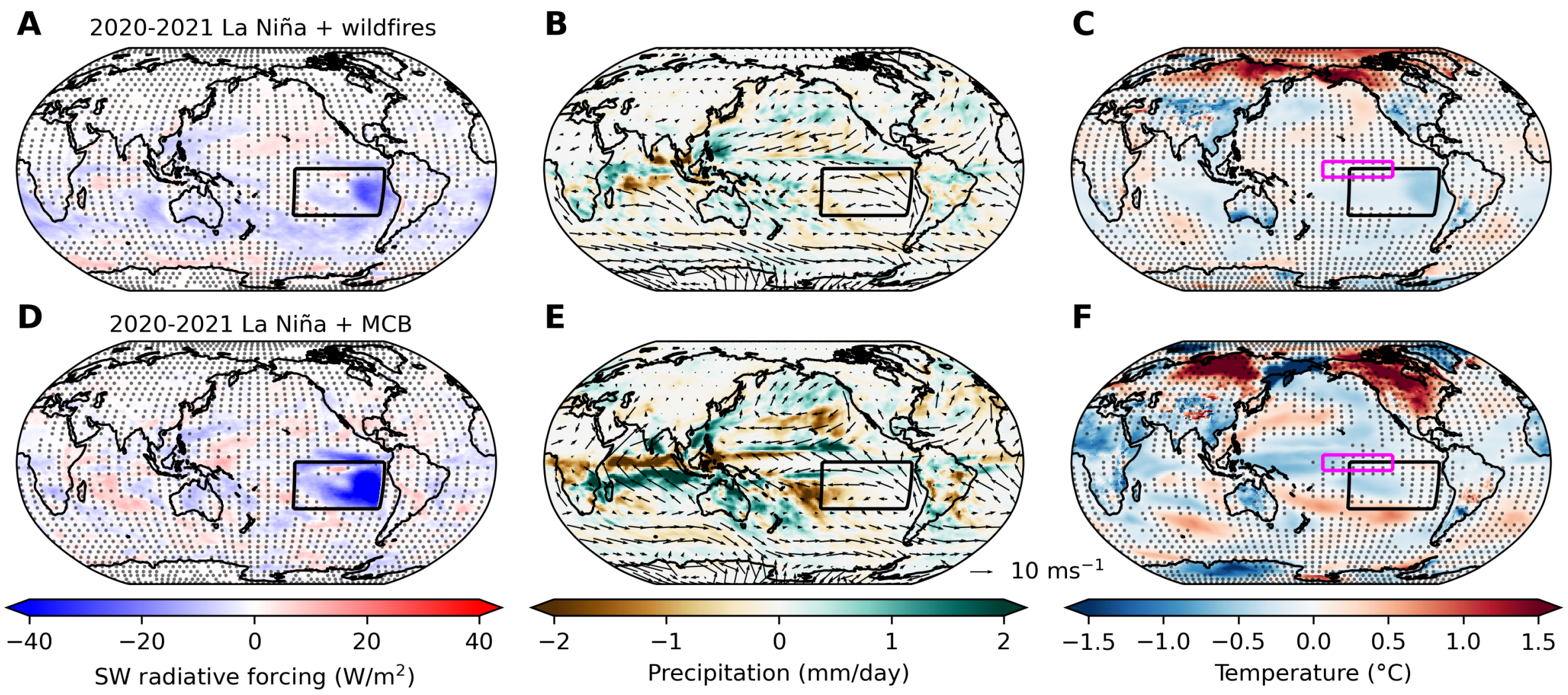

**Fig. 1. Comparison of responses to the Australian wildfires and MCB preceding the 2020-2021 La Niña.** (**A, D**) Mean shortwave cloud forcing response averaged between January to March 2020. (**B, E**) Mean precipitation response overlaid with mean wind near-surface wind vectors averaged between December 2020 to February 2021. (**C, F**) Mean surface temperature response averaged between December 2020 to February 2021 for the Australian wildfires (A-C) and MCB (D-F). Model output for the Australian wildfires is from the August initialized 20-member AUFIRE SMYLE experiment (*28*). MCB is applied in the black box region from January to March prior to the peak La Niña event. The magenta box (C, F) indicates the Niño3.4 region (120 °W to 170 °W, 5 °S to 5 °N). Stippling indicates insignificant anomalies less than two times the standard error of the mean of the SMYLE control.

**Deliberate marine cloud brightening to modify El Niño**

While implementing MCB in the SESP region can reproduce the La Niña response from the 2019-2020 Australian wildfires, the timing of MCB to effectively modify ENSO would not correspond to the seasonal cycle of wildfires. Physical limitations in ENSO forecasting, in particular the spring predictability barrier (*33*), would be a determining constraint on MCB initiation because models have low skill in predicting ENSO events before boreal summer. Additionally, the duration of MCB deployment could vary depending on the desired climate outcome and available resources. We simulate a set of strategies varying MCB initiation and termination preceding the 2015-2016 El Niño, one of the strongest events in the 21st century (*34*), to characterize the importance of timing in modifying El Niño.

MCB deployed after the spring predictability barrier dampens the 2015-2016 El Niño, and the responses depend on the timing of intervention (Fig. 2A; fig. S1). During peak El Niño (Fig. 2B; "ENSO peak" defined here as DJF), the earliest and longest "Full effort" MCB (Fig. 2C; June-February MCB) results in the strongest cooling in the SESP brightening region (-1.18 °C) and reductions in Niño3.4 SST (-1.48 °C), virtually restoring neutral ENSO conditions by the end of the event peak. On the other end, the latest and shortest "11th hour" MCB (December-February MCB) leads to less cooling in the SESP (-0.52 °C) and the smallest Niño3.4 SST reductions (-0.31 °C). "Early action" MCB (June-August MCB) has the same initiation month as the Full effort strategy and the same duration as the 11th hour strategy yet causes the least amount of cooling in the SESP (-0.31 °C) and moderate Niño3.4 SST cooling (-0.83 °C).

Considering ENSO indicators beyond SST, the three scenarios with the longest durations have the greatest efficacy in altering atmospheric conditions, while the earliest initiating scenarios produce the largest changes in the oceanic state during the event peak (table S2). Perhaps intuitively, these results demonstrate that starting MCB early and leaving it on for the longest duration is more effective at weakening El Niño than implementing MCB only during the event peak. We find similar responses for the 1997-1998 El Niño (fig. S2), an event characterized by strong eastern Pacific warming compared to the 2015-2016 El Niño which was a mixed regime with central Pacific warming (*34*), suggesting these findings are robust to different types of El Niño events within CESM2.

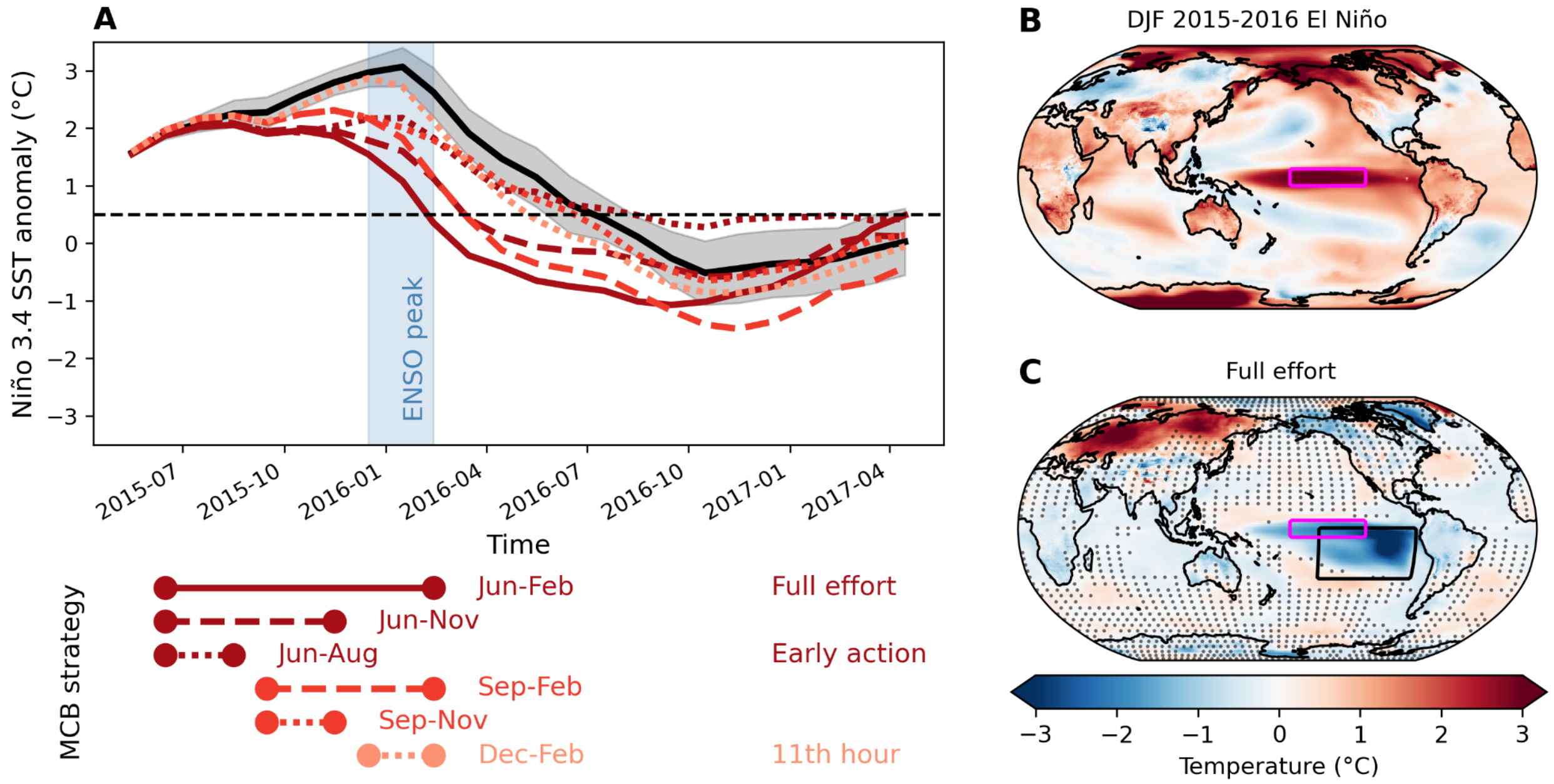

**Fig. 2. Surface temperature response to MCB during the 2015-2016 El Niño.** (**A**) Ensemble mean Niño 3.4 SST anomaly time series for the 2015-2016 El Niño. Black line shows the SMYLE control ensemble mean (with two standard error shading) and the red/orange lines show the SMYLE MCB ensemble means for different initiation months and durations. Dashed black line indicates the +0.5 °C Niño3.4 threshold above which SST anomalies are characterized as El Niño. (**B**) Ensemble mean DJF surface temperature anomalies for control SMYLE relative to the historical SMYLE monthly climatology (1970-2014). (**C**) Ensemble mean DJF surface temperature response to Full effort MCB. Stippling indicates insignificant MCB anomalies less than two times the standard error of the mean of the control SMYLE. Boxes indicated the MCB seeding (black) and Niño 3.4 region (magenta).

We find that MCB also changes the timing and magnitude of the ensuing La Niña event following El Niño. All MCB strategies result in colder Niño3.4 SST anomalies in boreal fall after the El Niño peak, except Early action MCB which stabilizes around neutral ENSO SSTs through the next year (Fig. 2A). The three longest duration MCB strategies tend towards negative Niño3.4 SST anomalies of -0.5 °C or less (that is, a La Niña state) seasons earlier than without MCB. Changes to the ensuing La Niña are less pronounced or insignificant following the 1997-1998 El Niño (fig. S2A), indicating that La Niña effects are sensitive to the preceding El Niño conditions.

## Early marine cloud brightening disrupts ENSO growth feedbacks

If El Niño dampening was strictly caused by the cloud forcing imposed by MCB, we would expect the MCB strategies with the greatest radiative perturbations to have the strongest reductions in peak Niño3.4 SST anomalies. However, this relationship is only true for Full effort MCB, which has the largest mean shortwave forcing response from June to February in the seeding region (-27.1 W/m$^2$; fig. S3) and DJF Niño3.4 SST anomaly (-1.88 °C; fig. S4). The other five strategies differ in ordering between their mean forcing and SST responses, with the largest discrepancy for 11th hour MCB which has the smallest peak Niño3.4 SST anomaly (-0.31 °C) but only the third smallest mean forcing (-8.55 W/m$^2$) likely due to the stronger incoming solar radiation during boreal winter in the Southern Hemisphere. Thus, the pathway for MCB to modify ENSO cannot be fully explained by the atmospheric thermodynamic response to radiative perturbations. Instead, MCB must also induce dynamical changes between the atmosphere and ocean that feed back onto the ENSO cycle.

To diagnose the dynamical responses to MCB, we delve into the progression of physical processes that drive El Niño growth and decay. In boreal spring, downwelling Kelvin waves deepen the thermocline, reduce upwelling, and lead to warmer SSTs in the eastern tropical Pacific, including in the Niño3.4 region. This weakens the characteristic zonal SST gradient. Late boreal summer through fall marks the typical growth phase of El Niño during which positive Bjerknes feedbacks (*35*) amplify the weakened zonal SST gradient and easterly trade winds in the tropical Pacific (*36*).

By boreal winter, an unperturbed mature El Niño is typically associated with decreased Walker cell strength (see 'Walker circulation strength index', in Materials and Methods), a more uniform equatorial SST gradient, and a shoaled thermocline slope index (see 'Thermocline slope index', in Materials and Methods). However, MCB results in a strengthened Walker cell, increased zonal SST gradient, and a steeper thermocline slope compared to the typical El Niño responses, effectively restoring neutral ENSO conditions for the atmosphere and ocean in the eastern Pacific, while having a more moderate effect in the western Pacific (Fig. 3).

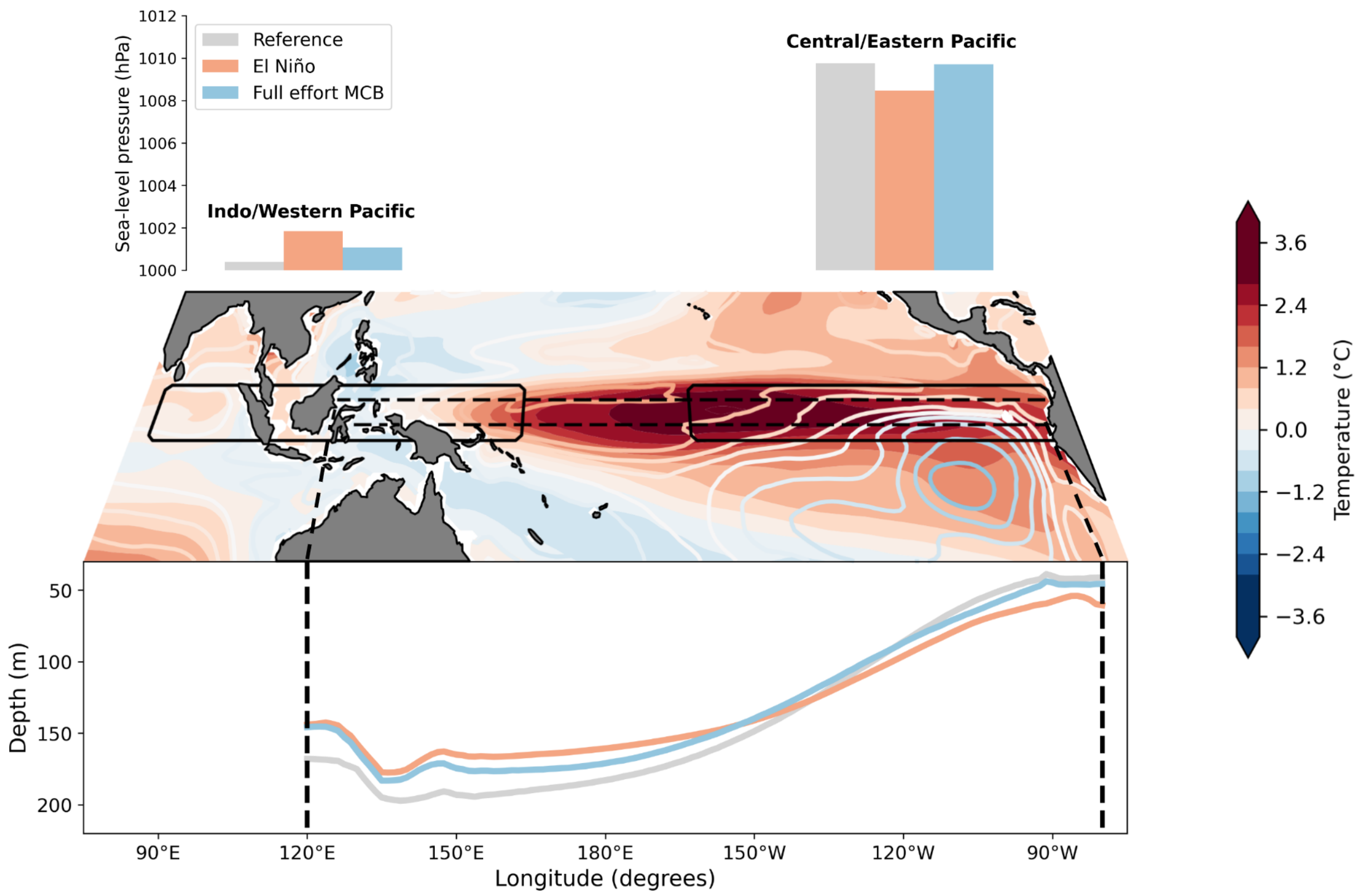

**Fig. 3. Mechanisms of El Niño and marine cloud brightening**. Shaded contours on map show the ensemble mean DJF SST anomalies for the control 2015-2016 El Niño relative to the reference monthly climatology (1970-2014). Colored contour lines on map show the ensemble mean DJF SST anomalies due to Full effort MCB relative to the reference climatology. The ensemble mean

DJF sea-level pressure (SLP) (top; bars) and 20 °C isotherm (Z20) (bottom; lines) are plotted for the reference climatology (gray), control 2015-2016 El Niño (red) and Full effort MCB during the 2015-2016 El Niño (blue).

This change in typical atmospheric and oceanic conditions associated with a neutral ENSO state depends on the timing of MCB deployment. Both Walker cell strength and thermocline slope index (fig. S5) are statistically unchanged by 11th hour MCB. Because this strategy initiates MCB after the El Niño growth phase, at this stage, MCB predominantly imposes local radiative cooling in the SESP region rather than alters the dynamics of the ENSO cycle. All other MCB strategies significantly increase Walker cell strength, steepen thermocline slope, and reduce Niño3.4 SST anomalies during the El Niño peak. This suggests that MCB implemented in the SESP during the growth phase dampens El Niño most effectively because it disrupts the nonlinear Bjerknes feedbacks that would normally enhance El Niño conditions.

## Regional climate teleconnections to marine cloud brightening and El Niño

While the societal impacts of El Niño events are heterogeneous, El Niño has been linked to net reduced economic growth (*7*). El Niño years are also warmer at the global mean level (*37–39*) so, on average, will boost the predominant effects of anthropogenic warming (*2*). While our analysis illustrates that MCB can dampen El Niño as measured by the Niño3.4 index (Fig. 2A), Southern Oscillation Index (table S2), and some of its physical manifestations across the tropical Pacific (Fig. 3), whether MCB is ultimately beneficial will depend on whether the harmful regional impacts associated with El Niño are reduced.

In CESM2, five major historical El Niño events capture many of the typical regional temperature and precipitation changes associated with El Niño (fig. S6; see 'El Niño teleconnection regions', in Materials and Methods). While Full effort MCB ameliorates teleconnections in all regions that experience the largest changes during peak El Niño, 11th hour MCB exacerbates El Niño impacts for nearly two thirds of the regions (Fig. 4). Early action MCB generally results in responses that fall between the Full effort and 11th hour strategies with a notable, but statistically insignificant, exception in east Brazil (Fig. 4H). Similar responses occur under the 1997-1998 El Niño (fig. S7).

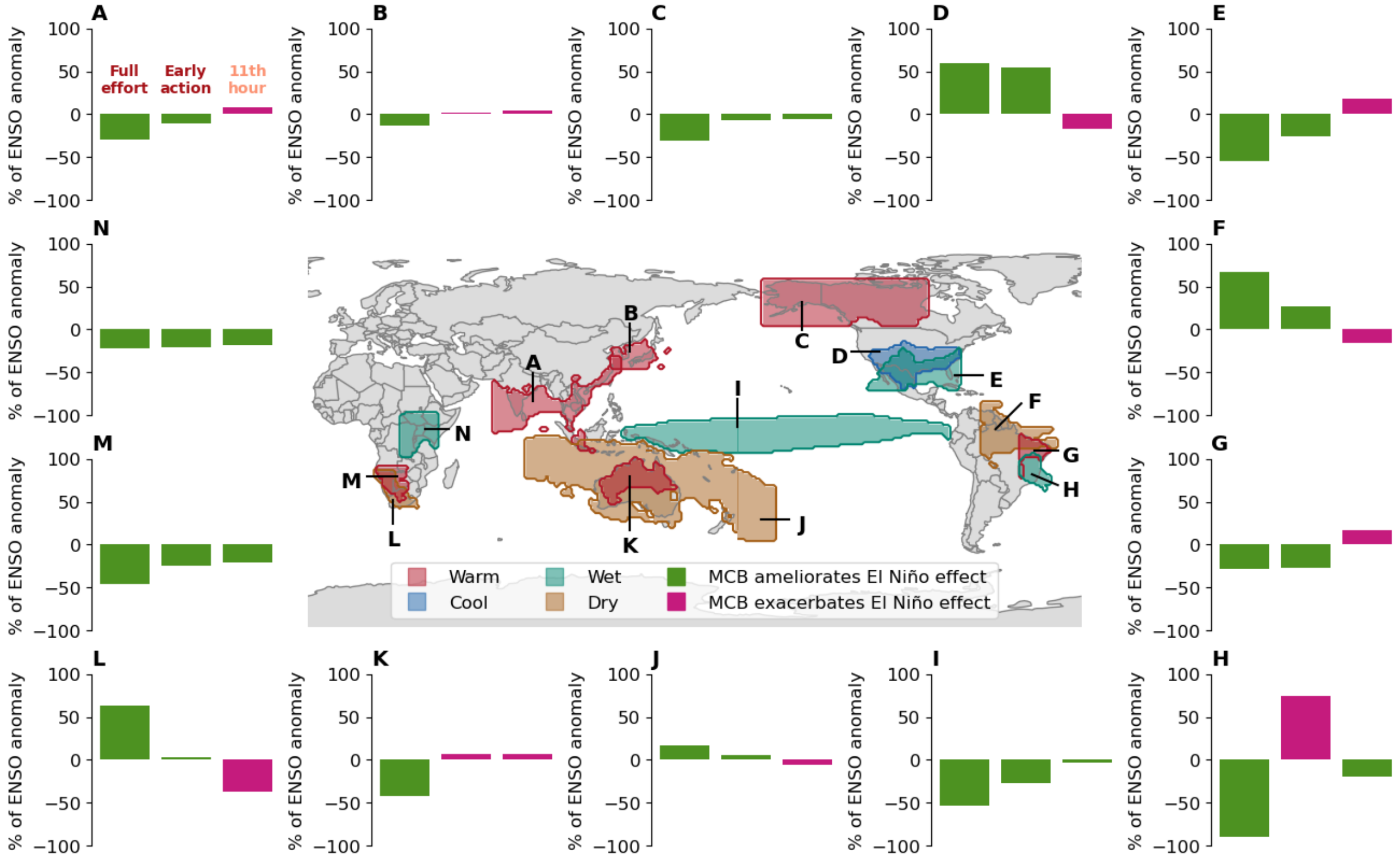

**Fig. 4. Change in DJF climate teleconnections to MCB compared to the 2015-2016 El Niño anomalies for select regions.** The left bars show the ensemble mean percent change of the control 2015-2016 El Niño climate teleconnection due to Full effort MCB (June-February), the middle bars show responses from Early action MCB (June-August), and the right bars show the response due to 11th hour MCB (December-February). The regional responses are (**A**) South Asia warming, (**B**) Japan warming, (**C**) Boreal North America warming, (**D**) Southeast United States cooling, (**E**) South United States wettening, (**F**), Northeast South America drying, (**G**) East Brazil warming, (**H**), East Brazil wettening, (**I**) Equatorial Pacific wettening, (**J**) Oceania drying, (**K**) North Australia warming, (**L**) South Africa drying, (**M**) South Africa warming, (**N**) East Africa wettening.

While the results are consistent with our physical expectations, all regional changes are statistically insignificant in our 10-member ensembles, except for reductions in drying in northeast South America (Fig. 4F) and warming in southern Africa (Fig. 4M) under Full effort MCB, reflecting the regional internal variability even using a seasonal initialized prediction system. Nonetheless, these results demonstrate that timely strategic MCB deployment could ameliorate remote teleconnections caused by El Niño.

## Climate variability as a novel target for solar geoengineering research

In this study, we demonstrate the possibility that targeted marine cloud brightening could modify ENSO and reduce its associated remote climate effects. By exploiting a unique opportunistic experiment provided by the 2019-2020 Australian wildfires, we show that both the observed cloud brightening and ensuing La Niña-like response can be reproduced by simulating MCB in the southeast Pacific. Given current ENSO forecasting limitations, strategic deployments of MCB to dampen El Niño would need to be deployed after the spring predictability barrier. MCB implemented any time during the boreal summer and fall growth phase effectively ameliorates many of the remote teleconnections that drive El Niño damages.

Best available estimates suggest that a nine-month deployment of approximately 2,000 merchant class ships in the SESP region would be sufficient to increase cloud droplet number concentrations to approximately the levels simulated in our experiments (*31*). Applying recent operating cost estimates (*40*), such an operation would cost on the order of billions of dollars, compared to the trillions of dollars of economic damages associated with large historical El Niño events (*7*), suggesting a favorable cost benefit ratio, even if SESP MCB only partially corrects for remote climate effects of such an event.

While the seasonal initialized prediction ensemble used here represents an improvement on the tools used in previous MCB modeling work, it uses only one Earth system model and is therefore subject to the biases and limitations of single model studies. Our analysis focuses on two illustrative historical El Niño events, but future work would benefit from testing the robustness of these results against a broader diversity of historical and projected ENSO conditions, including any long-term impacts on the climate system and changes to the ensuing La Niña, such as shifts in the probability of multi-year La Niña events. It is plausible that short-term MCB not only influences the immediate ENSO event but also changes future ENSO magnitude and frequency or even the background mean state.

Solar geoengineering, while initially conceived as a drastic approach to reduce warming from greenhouse gases, could be a promising way to modify climate variability by targeting seasonal events including El Niño. As long-term anthropogenic warming and short-term natural variability compound to produce extreme weather events, our study suggests it may be worth considering whether targeting natural variability, rather than the forced response to greenhouse gases, could result in similar risk reduction for a smaller or shorter duration intervention in the climate system.

**Acknowledgments:** The simulations conducted are possible thanks to the high-performance computing support from Cheyenne and Derecho provided by the National Science Foundation (NSF) National Center for Atmospheric Research (NCAR) Computational Information Systems Laboratory.

**Funding:**

NCAR Early Career Faculty Innovator Program Cooperative Agreement No. 1755088 (JSW, KR)

NCAR Cooperative Agreement No. 1852977 (JTF, NR, CCC)

National Defense Science and Engineering Graduate Fellowship Program (JSW)

Achievement Rewards for College Scientists Foundation (JSW)

NASA Awards 80NSSC21K1191, 80NSSC17K0565, and 80NSSC22K0046 (JTF)

NSF Award 2103843 (JTF)

U.S. Department of Energy, Office of Science, Office of Biological & Environmental Research (BER), Regional and Global Model Analysis (RGMA) component of the Earth and Environmental System Modeling Program Award Number DE-SC0022070 (NR)

NOAA ERB grant NA22OAR4310481 (CCC)

**Author contributions:** JTF and KR conceived of the study. JSW, JTF, and KR designed the experiments and developed the methodology. JSW, NR, and CCC developed CESM code modifications. NR ran the simulations. JSW and KR developed visualizations of the results. JSW analyzed the output and wrote the original manuscript draft. All co-authors contributed to review and editing.

**Competing interests:** Authors declare that they have no competing interests.

**Data and materials availability:** All data, code, and materials used in the analysis will be made available at an online repository following submission.

**Supplementary Materials**

Materials and Methods

Figs. S1 to S7

Tables S1 to S2

References (*41–44*)

## Materials and Methods

Simulating MCB with the Seasonal-to-Multiyear Large Ensemble using CESM2

We use the Seasonal-to-Multiyear Large Ensemble (SMYLE) (*29*) using the Community Earth System Model version 2 (CESM2) (*27*) to simulate the climate responses of MCB in the SESP region. SMYLE is an initialized prediction system designed to test prediction skill for lead times ranging from one month to two years (*29*). Following a similar model setup as used in Fasullo et al. (2023), we run two-year-long MCB simulations for ten ensemble members preceding three historical ENSO events: (1) the 2020-2021 La Niña in order to validate the observational analogue presented in Fasullo et al. (2023) and illustrate the connection to MCB, (2) the 2015-2016 El Niño to test MCB's potential efficacy in weakening the most extreme El Niño of the 21$^{st}$ century (*34*), and (3) the 1997-1998 El Niño to check the robustness of (2). We use the corresponding 10 member hindcast simulations from the SMYLE experiment with no MCB as our control to compute the responses to MCB and the full 20 member SMYLE experiment with no MCB for our reference historical monthly climatology (1970-2014).

MCB perturbations are represented in CESM2 following the same methodology as in Wan et al. (2024) by nudging the cloud droplet number concentration to a constant of 500 particles/cm$^3$ over the target SESP region (Lat: 30 °S to 0 °; Lon: 150 °W to 85 °W). In our first set of experiments for (1), we initialize each member in November 2019 and apply MCB continuously in the SESP region from January through March 2020, then shut MCB off for the remainder of the simulation to allow for the development of any ENSO conditions. In (2) and (3), we run a set of six May initialized experiments (10 ensemble members each) with varying MCB initiation months and durations. Boreal spring initialization was chosen to reduce the simulation runtime prior to MCB initiation. The six MCB strategies are implemented as varied combinations of seasonal (JJA, SON, and DJF) deployments with the earliest experiments initiating MCB in June and the latest experiments ending MCB in February (as detailed in Table S1).

Walker circulation strength index

We use Walker cell strength as a proxy for the atmospheric component of the Bjerknes feedback (*35*) due to the tight coupling between the Walker circulation and equatorial Pacific SSTs. The Walker circulation strength index (*41*, *42*) is defined as the difference in sea-level pressure (SLP) between the Indian Ocean/west Pacific (80 °E to 160 °E, 5 °S to 5 °N) and central/eastern Pacific (160 °W to 80 °W, 5 °S to 5 °N) regions (Fig. 3; fig. S5A). A high Walker circulation strength index amplifies positive Bjerknes feedbacks while a low Walker strength index weakens Bjerknes feedbacks.

Thermocline slope index

We approximate the oceanic component of the Bjerknes feedback using the thermocline depth approximated as the depth of the ensemble mean DJF 20 °C isotherm (Z20) over the equatorial Pacific (120 °E to 80 °W, 2 °S to 2 °N) following Timmerman et al. (2018) (Fig. 3). We calculate the thermocline slope index (fig. S5B) as the difference in Z20 anomalies (relative to the reference SMYLE climatology) between the western (160 °E to 150 °W, 2 °S to 2 °N) and eastern Pacific (90 °W to 140 °W, 2 °S to 2 °N) (*43*). A steeper thermocline slope is associated with neutral ENSO/La Niña conditions while a shoaled thermocline is indicative of El Niño conditions.

El Niño teleconnection regions

Similar to the regional impacts forecasted by the U.S. National Oceanic and Atmospheric Administration (*44*) (NOAA), we define a set of teleconnection regions (Fig. 4). We compute the DJF composite surface temperature and precipitation anomalies for major historical El Niño events (1972, 1982, 1991, 1997, and 2015) from the SMYLE reference simulations and select pixels greater than 0.2 times the reference standard deviation (fig. S6). We choose this 0.2 standard deviation threshold for both temperature and precipitation to capture the key impact regions identified by NOAA using the SMYLE reference simulation output. For each region, we then calculate the change in temperature or precipitation due to MCB (relative to the control El Niño) and normalize by the control El Niño anomaly (relative to the reference period 1970-2014) to obtain the percent change due to MCB relative to the original El Niño event.

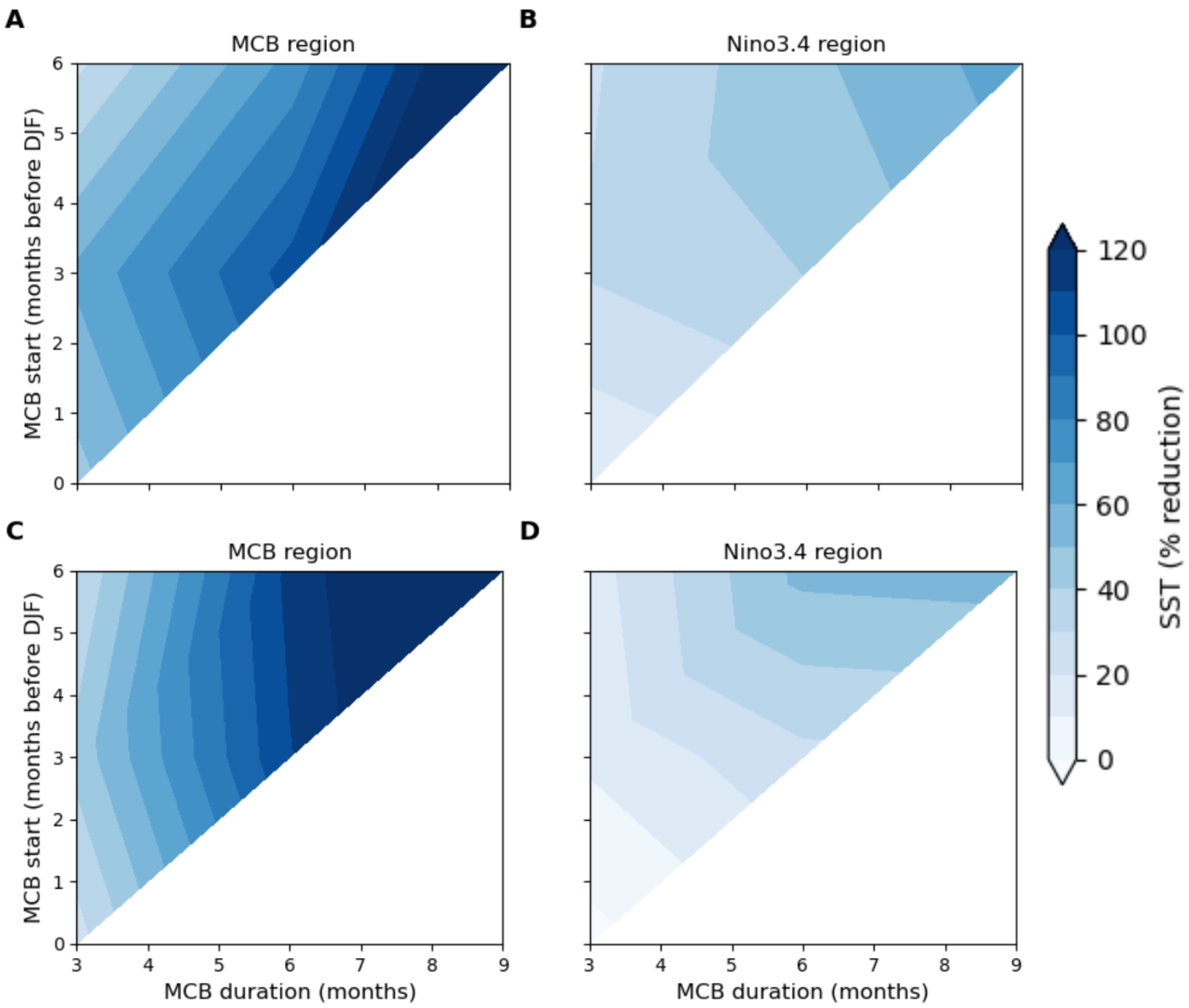

**Fig. S1.**

Contour plots of the percent reduction in DJF sea surface temperatures due to varying MCB initiation months and durations. (**A, B**) DJF SST reductions relative to the no MCB control for the 2015-2016 El Niño experiments in the MCB region (A) and Niño3.4 region (B). (**C, D**) Same as (A, B) for the 1997-1998 El Niño experiments.

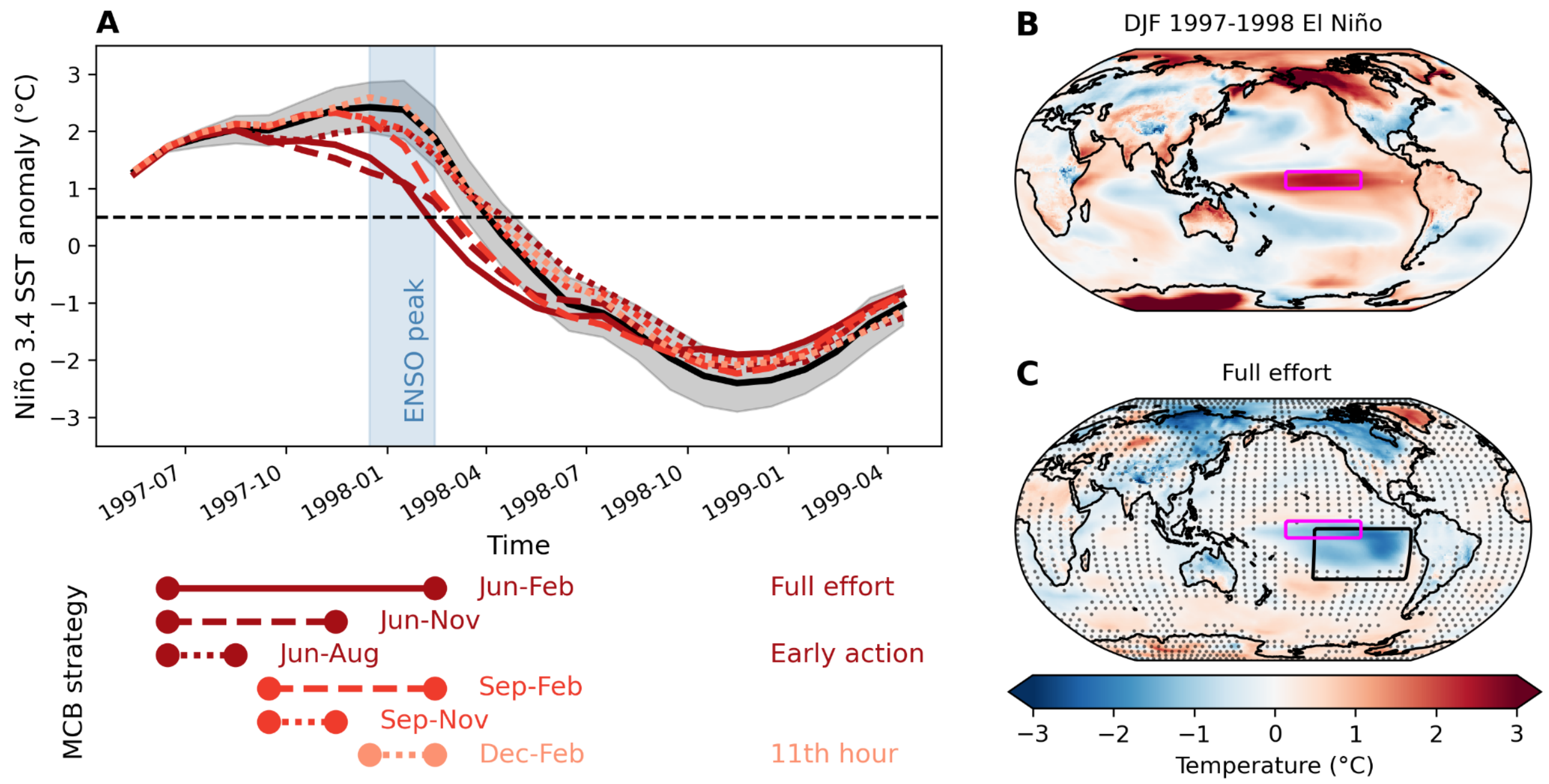

**Fig. S2.**

Surface temperature response to MCB during the 1997-1998 El Niño. (**A**) Ensemble mean Niño 3.4 SST anomaly time series for the 1997-1998 El Niño. Black line shows the SMYLE control ensemble mean (with two standard error shading) and the red/orange lines show the SMYLE MCB ensemble means for different initiation months and durations. Dashed black line indicates the +0.5 °C Niño3.4 threshold above which SST anomalies are characterized as El Niño. (**B**) Ensemble mean DJF surface temperature anomalies for control SMYLE relative to the historical SMYLE monthly climatology (1970-2014). (**C**) Ensemble mean DJF surface temperature response to Full effort MCB. Stippling indicates insignificant MCB anomalies less than two times the standard error of the mean of the control SMYLE. Boxes indicated the MCB seeding (black) and Niño 3.4 region (magenta).

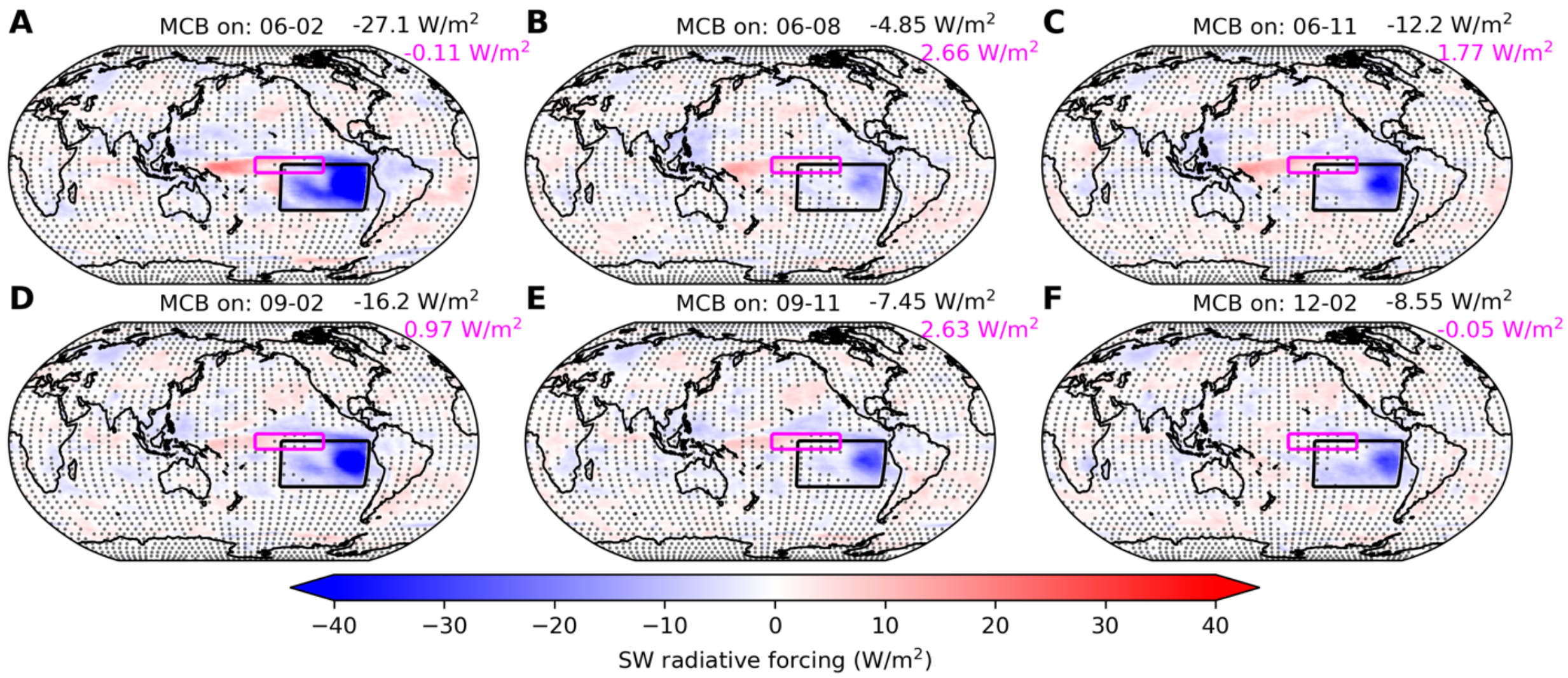

**Fig. S3.**

Shortwave cloud forcing response to each MCB strategy during the 2015-2016 El Niño. Ensemble mean shortwave cloud forcing response due to MCB from June to February for (**A**) Full effort (June-February) (**B**) Early action (June-August) (**C**) June-November (**D**) September-February (**E**) September-November and (**F**) 11th hour (December-February) MCB strategies. Stippling indicates insignificant MCB anomalies less than two times the standard error of the mean of the control SMYLE. Top right values show the area-weighted mean over the MCB seeding region (black) and Niño 3.4 region (magenta).

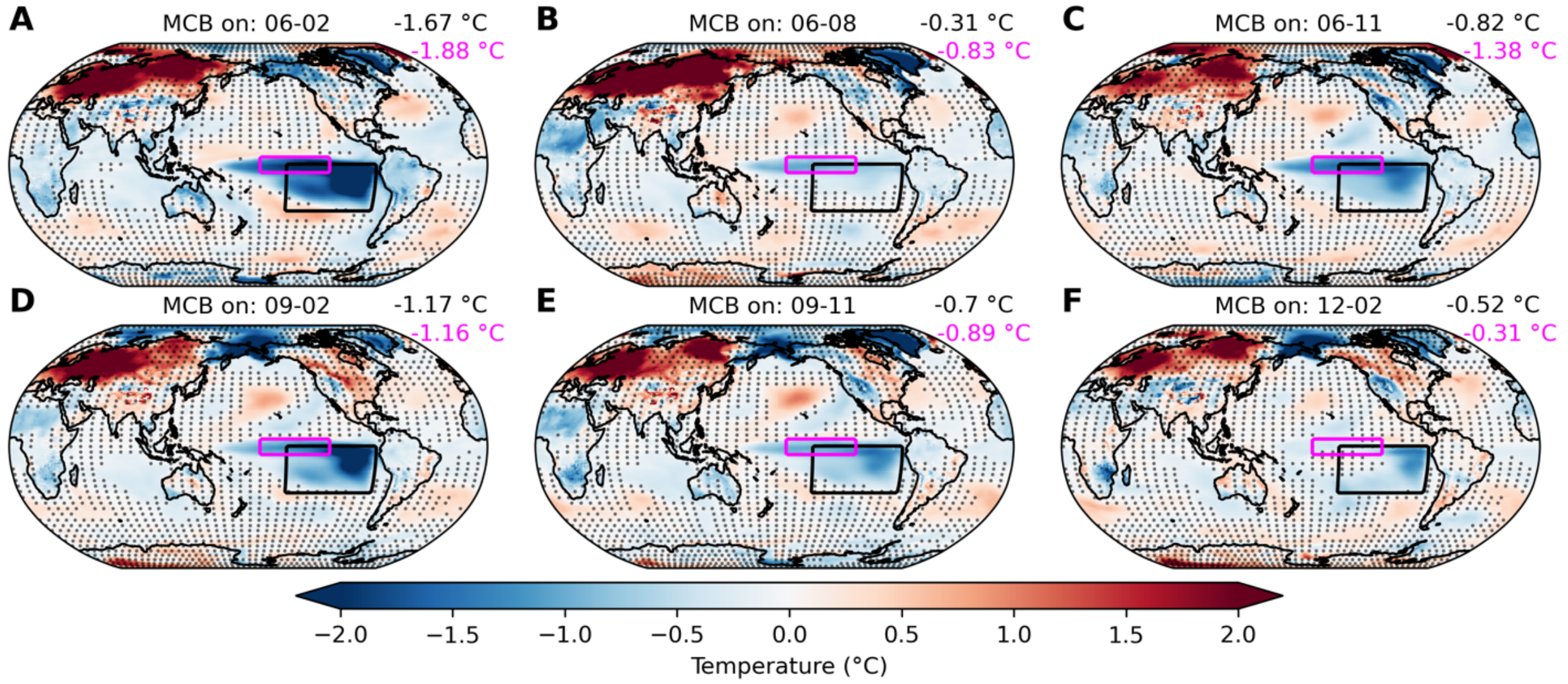

**Fig. S4.**

DJF surface temperature response to each MCB strategy during the 2015-2016 El Niño. Ensemble mean surface temperature response due to MCB during the El Niño peak (DJF) for (**A**) Full effort (June-February) (**B**) Early action (June-August) (**C**) June-November (**D**) September-February (**E**) September-November and (**F**) 11th hour (December-February) MCB strategies. Stippling indicates insignificant MCB anomalies less than two times the standard error of the mean of the control SMYLE. Top right values show the area-weighted mean over the MCB seeding region (black) and Niño 3.4 region (magenta).

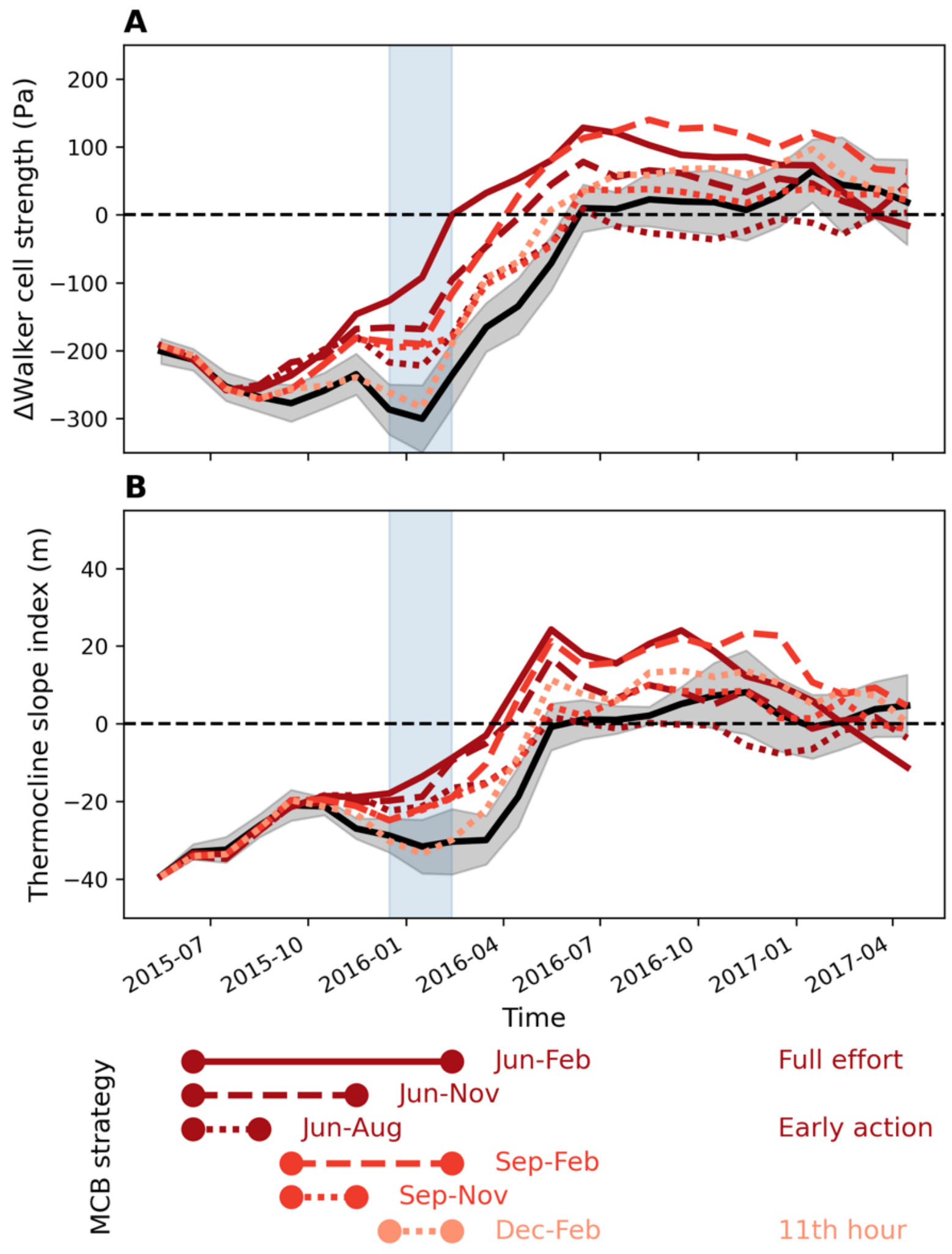

**Fig. S5.**

Atmosphere and ocean El Niño growth feedback mechanism responses to MCB. (**A**) Ensemble mean Walker cell strength anomaly (see 'Walker circulation strength index' in Methods) for MCB and the control 2015-2016 El Niño relative to the reference SMYLE climatology (1970-2014). (**B**) Ensemble mean thermocline slope index anomaly (see 'Thermocline slope index' in Methods) for MCB and the control 2015-2016 El Niño relative to the reference climatology.

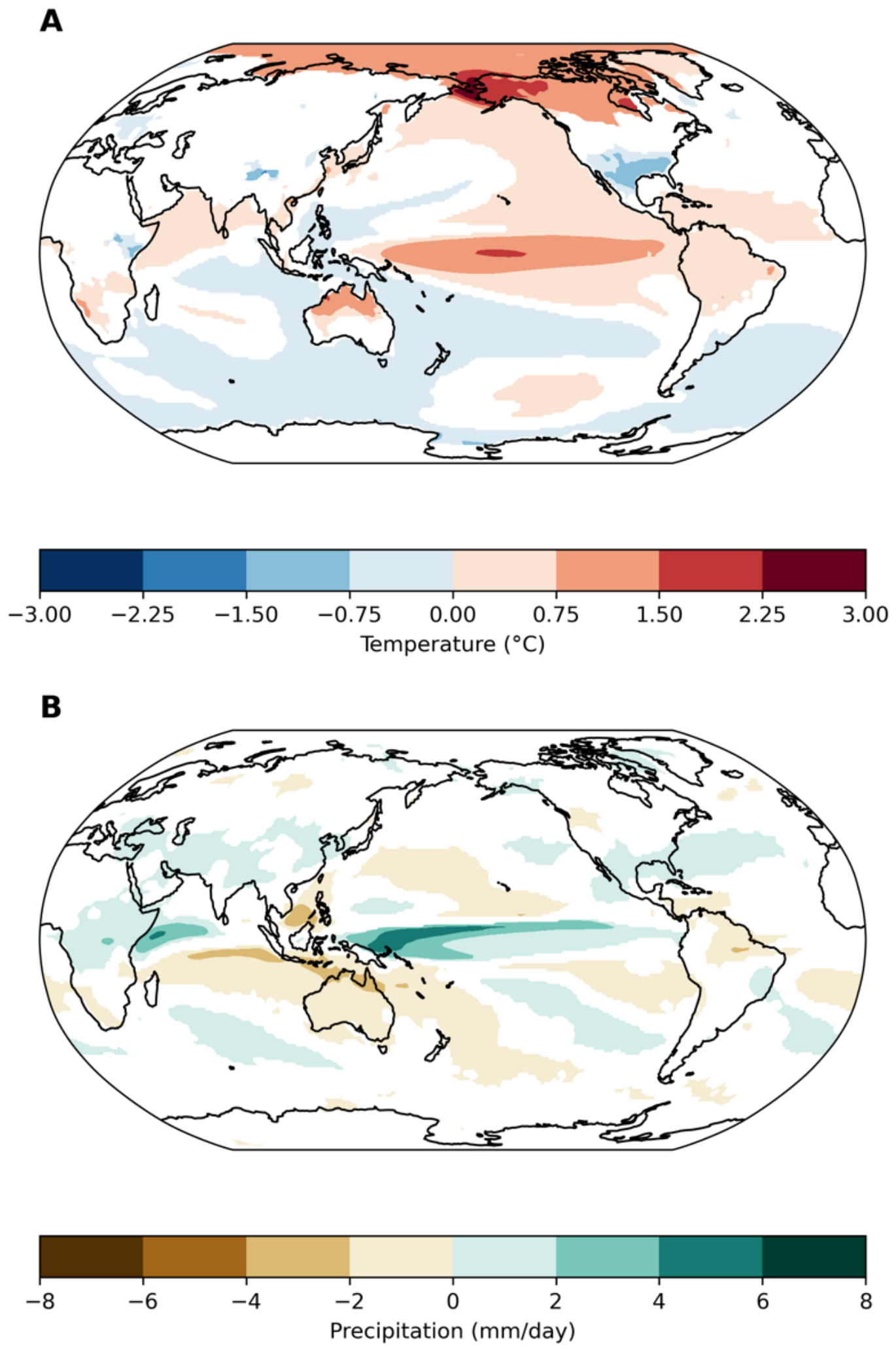

**Fig. S6.**
DJF composite surface temperature and precipitation anomalies for major historical El Niño events. Selected major historical El Niño events from the SMYLE reference simulations include the 1972, 1982, 1991, 1997, and 2015 events. Displayed pixels are greater than 0.2 times the reference standard deviation for (**A**) surface temperature and (**B**) precipitation.

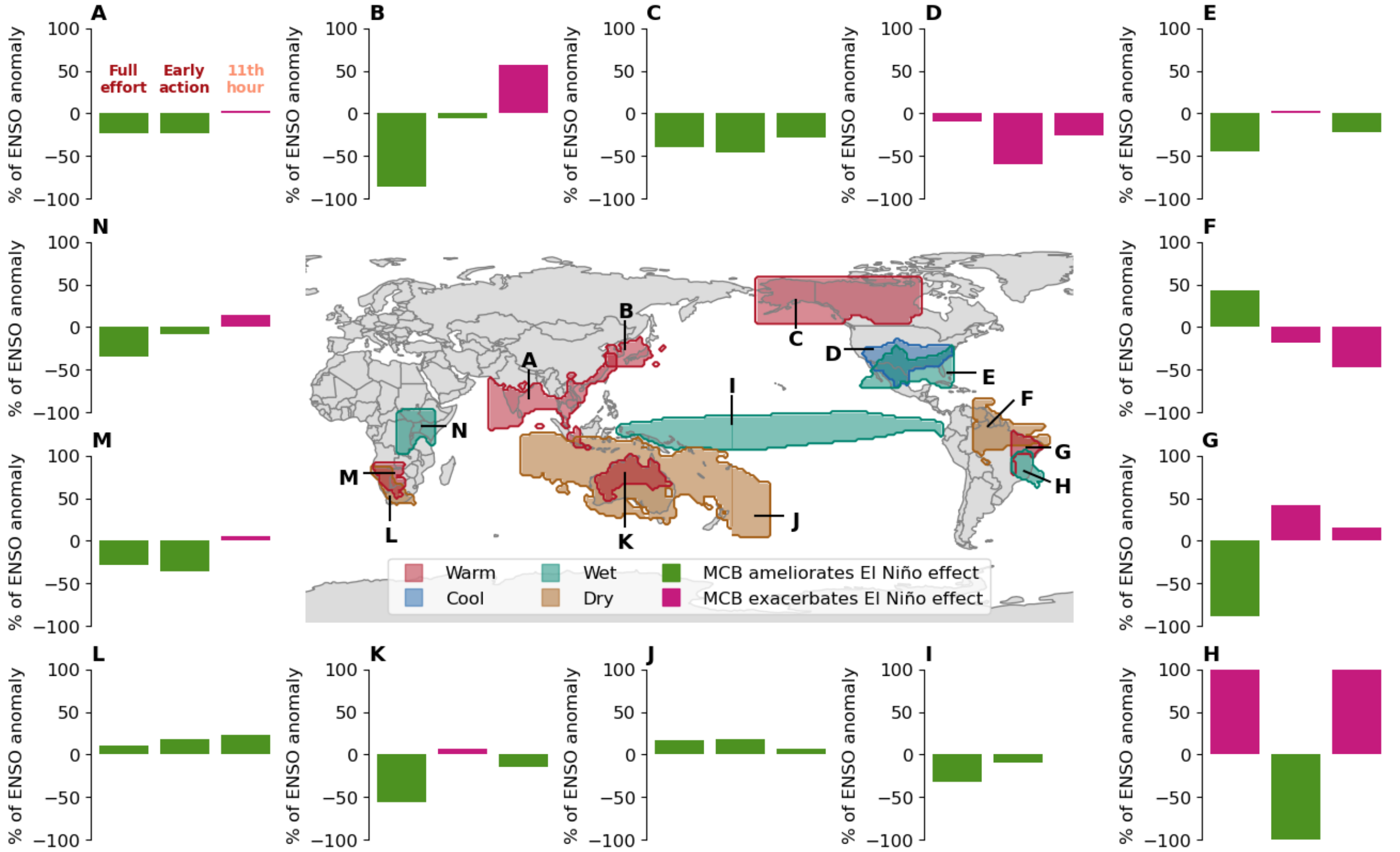

**Fig. S7.**
Change in DJF climate teleconnections to MCB compared to the 1997-1998 El Niño anomalies for select regions. The left bars show the ensemble mean percent change of the control 1997-1998 El Niño climate teleconnection due to Full effort MCB (June-February), the middle bars show responses from Early action MCB (June-August), and the right bars show the response due to 11th hour MCB (December-February). The regional responses are (**A**) South Asia warming, (**B**) Japan warming, (**C**) Boreal North America warming, (**D**) Southeast United States cooling, (**E**) South United States wettening, (**F**), Northeast South America drying, (**G**) East Brazil warming, (**H**), East Brazil wettening, (**I**) Equatorial Pacific wettening, (**J**) Oceania drying, (**K**) North Australia warming, (**L**) South Africa drying, (**M**) South Africa warming, (**N**) East Africa wettening.

**Table S1.**

SMYLE-MCB simulations for (1) 2020-2021 La Niña, (2) 2015-2016 El Niño, and (3) 1997-1998 El Niño. All MCB simulations have a corresponding control member from the SMYLE ensemble with no MCB (not shown).

| Experiment set | MCB strategy (if applicable) | Simulation long name | Init. month | Simulation duration | MCB period |
|---|---|---|---|---|---|
| 1. | 2020-2021 La Niña + MCB | b.e21.BSMYLE.f09_g17.MCB.2019-11.0* | Nov | 2019-11 to 2021-10 | 2020-01 to 2020-03 |
| 2. | Early action | b.e21.BSMYLE.f09_g17.MCB_2015-05_06-08.2015-05.0* | May | 2015-05 to 2017-04 | 2015-06 to 2015-08 |
| 2. | | b.e21.BSMYLE.f09_g17.MCB_2015-05_06-11.2015-05.0* | May | 2015-05 to 2017-04 | 2015-06 to 2015-11 |
| 2. | Full effort | b.e21.BSMYLE.f09_g17.MCB_2015-05_06–02.2015-05.0* | May | 2015-05 to 2017-04 | 2015-06 to 2016-02 |
| 2. | | b.e21.BSMYLE.f09_g17.MCB_2015-05_09–11.2015-05.0* | May | 2015-05 to 2017-04 | 2015-09 to 2015-11 |
| 2. | | b.e21.BSMYLE.f09_g17.MCB_2015-05_09–02.2015-05.0* | May | 2015-05 to 2017-04 | 2015-09 to 2016-02 |
| 2. | 11th hour | b.e21.BSMYLE.f09_g17.MCB_2015-05_12–02.2015-05.0* | May | 2015-05 to 2017-04 | 2015-12 to 2016-02 |
| 3. | Early action | b.e21.BSMYLE.f09_g17.MCB_1997-05_06-08.1997-05.0* | May | 1997-05 to 1999-04 | 1997-06 to 1997-08 |
| 3. | | b.e21.BSMYLE.f09_g17.MCB_1997-05_06-11.1997-05.0* | May | 1997-05 to 1999-04 | 1997-06 to 1997-11 |
| 3. | Full effort | b.e21.BSMYLE.f09_g17.MCB_1997-05_06–02.1997-05.0* | May | 1997-05 to 1999-04 | 1997-06 to 1998-02 |
| 3. | | b.e21.BSMYLE.f09_g17.MCB_1997-05_09–11.1997-05.0* | May | 1997-05 to 1999-04 | 1997-09 to 1997-11 |
| 3. | | b.e21.BSMYLE.f09_g17.MCB_1997-05_09–02.1997-05.0* | May | 1997-05 to 1999-04 | 1997-09 to 1998-02 |
| 3. | 11th hour | b.e21.BSMYLE.f09_g17.MCB_1997-05_12–02.1997-05.0* | May | 1997-05 to 1999-04 | 1997-12 to 1998-02 |

**Table S2.**

Summary of ENSO indicators during DJF of the 2015-2016 El Niño for six MCB strategies. The shaded cells illustrate each strategy's efficacy in weakening El Niño for a given ENSO indicator, where a darker blue indicates greater efficacy and a light blue indicates weaker efficacy.

| MCB strategy | Niño3.4 SST anomaly (°C) | Southern Oscillation Index | Walker strength index (Pa) | Thermocline slope index (m) |
|---|---|---|---|---|
| Full effort (Jun-Feb) | -1.88 | -0.01 | -72.9 | -13.5 |
| Jun-Nov | -1.38 | -0.54 | -143 | -16.0 |
| Early action (Jun-Aug) | -0.83 | -1.5 | -206 | -20.1 |
| Sep-Feb | -1.16 | -0.73 | -164 | -21.9 |
| Sep-Nov | -0.89 | -0.92 | -190 | -22.1 |
| 11th hour (Dec-Feb) | -0.31 | -2.07 | -245 | -31.3 |